\documentstyle[prb,aps]{revtex}
\tightenlines
\begin{document}
\draft

\title{Two-magnon Raman scattering in spin-ladder geometries
and the ratio of rung and leg exchange constants}
\author{P. J. Freitas and R. R. P. Singh}
\address{Department of Physics, One Shields Avenue, University of
California, Davis, California 95616-8677}
\date{July 14, 2000}

\maketitle


\begin{abstract}

We discuss ways in which the ratio of exchange constants
along the rungs and legs of a spin-ladder material
influences the two-magnon Raman scattering spectra and hence can
be determined from it. We show that within the Fleury-Loudon-Elliott
approach, the Raman line-shape does not change with polarization geometries.
This lineshape is well known to be difficult to calculate 
accurately from theory.
However, the Raman scattering intensities do vary with polarization
geometries, which are easy to calculate. 
With some assumptions about the Raman scattering
Hamiltonian, the latter can be used to estimate the ratio of exchange
constants. We apply these results to Sugai's recent measurements of 
Raman scattering from spin-ladder materials such as 
La$_6$Ca$_8$Cu$_{24}$O$_{41}$ and Sr$_{14}$Cu$_{24}$O$_{41}$.

\end{abstract}


\section{Introduction}
\label{sec:intro}

Determining the exchange constants and establishing the microscopic
spin Hamiltonian is an important step in understanding magnetic
properties of exotic new materials. In materials which have
predominantly a single exchange constant, the measurement of
uniform susceptibility and the determination of Curie-Weiss parameter
is sufficient to obtain the exchange constant. However, in recent years,
many complex materials have been synthesized, which have more than
one exchange constant. Furthermore, these exchange constants can be
so large that the Curie-Weiss regime may not be experimentally accessible.
In these cases alternative methods are needed to determine the
exchange parameters. Examples of such materials are cuprate
based spin-ladder materials, cousins of high temperature superconducting
materials, for which there is a substantial literature for determining
the exchange constants. The simplest methods for determining the
exchange constants involve measuring the temperature dependence
of uniform susceptibility, or Knight shift, or nuclear relaxation
rates, which can then be compared with detailed theoretical calculations
to obtain the exchange parameters. Such a fitting procedure is not very
accurate, as reflected in the range of values that exist in the 
literature for these materials.

Recently, Sugai \cite{Sugai1,Sugai2} noted that Raman scattering can be used to
determine the ratio of rung and leg exchange constants in spin-ladder 
materials. He argued that by varying the polarization direction
of incident and outgoing light, one might be able to shift the
Raman spectra in ways that can be related to the different exchange constants.
Our study is motivated by Sugai's work. However, we find that
the arguments used by Sugai to relate the position of the spectral
peaks to the different exchange constants in the spin-ladder materials
are incorrect. 
Within the Fleury-Loudon-Elliott \cite{Fleury&Loudon,Elliott} theory
the spectral lineshape does not change with polarization at all. 
This gives a simple explanation why Sugai always finds a very small shift
in the Raman spectra and sheds doubt on his inference that the
exchange constants in all the ladder materials are close to unity.

It is well known that an accurate calculation of Raman spectra for
low-dimensional spin-half antiferromagnets is very difficult due to
quantum fluctuations. \cite{Singh} And, short of a direct comparison
of the spectra with theory, it would 
seem that Raman spectra cannot be used to determine the ratio
of exchange constants. However, we show below that the Raman scattering
intensities do depend on polarization geometries in a way that
is easily calculated and related to the ratio of exchange
constants. Unfortunately, they come with an unknown factor, about which certain
assumptions need to be made before an estimate of the ratio 
of rung to leg exchange constants in the ladder material can be obtained.

\section{Spin-Ladder Heisenberg Models }
\label{sec:rectangular}

We begin with a system described by a Heisenberg model in ladder
geometry, with the Hamiltonian,
\begin{equation}
H = {J_r\sum_{\langle ij \rangle,r} {\vec S_{{\bf i}} \cdot  {\vec S_{{\bf j}}}}} + 
{J_l\sum_{\langle ij \rangle,l} {\vec S_{{\bf i}} \cdot  {\vec S_{{\bf j}}}}},
\label{1}
\end{equation}
where $J_r$ and $J_l$ denote the coupling constants for the rung and leg bonds 
of the ladder, respectively, and the sums are only over the bonds in the 
indicated directions. 
Our primary goal is to examine the dependence of Raman scattering on the ratio 
of exchange constants $J_r$ and $J_l$.

Within the Fleury-Loudon-Elliott approach, magnetic Raman scattering
is described by an effective Raman Hamiltonian or operator \cite{Parkinson}: 
\begin{equation}
{\cal H}_R = \sum_{\langle ij \rangle} J^\prime_{ij}(\hat \epsilon_{in}\cdot \hat r_{ij}) 
(\hat \epsilon_{out}\cdot \hat r_{ij})
{\vec S_{{\bf i}}} \cdot  {\vec S_{{\bf j}}},
\end{equation}
where the $\hat r_{ij}$ are unit vectors along the bond directions, and 
$\hat \epsilon_{in}$ and $\hat \epsilon_{out}$ are unit vectors indicating the
direction of polarization of the incident and scattered light, respectively. The
$J^\prime_{ij}$ are constants representing the strength of the Raman scattering
interaction between spins $i$ and $j$. In previous studies of the spin-ladder
geometry \cite{Sugai2,Natsume}, it has been assumed that $J_{ij}$ is 
nearest-neighbor and is a constant for all nearest-neighbor bonds. Thus it can 
be taken out of the summation. Such a constant simply sets the overall
scale for the scattering and does not influence any other result.
However, we believe that if the rung and leg exchange constants are
not equal, {\it a priori}, we cannot assume that the ratio of $J^\prime$ along
the rung and leg directions to be equal. Thus we proceed here with a more
general $J^\prime_{ij}$, and later consider possible scenarios for their
values, which would play an important role.

Following Sugai, it is most useful to consider the case where the incident and 
scattered light have parallel polarization directions, both lying in the plane 
of our 2D system. Thus, $\hat \epsilon_{in} = \hat \epsilon_{out}$. Most 
generally, we can denote the polarization with an angle $\theta$ with respect to 
the vertical bonds, which makes the effective Raman Hamiltonian
\begin{equation}
{\cal H}_R(\theta) = \cos^2\theta \sum_{\langle ij \rangle,r}J^\prime_{ij}
{{\vec S_{{\bf i}}} \cdot  {\vec S_{{\bf j}}}} + \sin^2\theta \sum_{\langle ij 
\rangle,l}J^\prime_{ij}{{\vec S_{{\bf i}}} \cdot {\vec S_{{\bf j}}}}.
\end{equation}
As a notational simplification, we define ${\cal H}_r = 
{\cal H}_R(0)$ and ${\cal H}_l = {\cal H}_R({\pi \over 2})$. Also, for reasons
of symmetry, we assume that all the $J^\prime_{ij}$ in each summation are the
same (we call them $J^\prime_r$ and $J^\prime_l$, respectively) so
\begin{equation}
{\cal H}_R(\theta) = J^\prime_r(\cos^2\theta){\cal H}_r + J^\prime_l
(\sin^2\theta){\cal H}_l.
\label{2}
\end{equation}
The two-magnon Raman scattering intensity as a function of frequency can be 
expressed using Fermi's golden rule,
\begin{equation}
I(\omega,\theta) = {\sum_n}^\prime {|\langle \psi_n|{\cal H}_R(\theta)|\psi_0
\rangle|^2 \delta(\omega-(E_n-E_0))},
\label{3}
\end{equation}
where $|\psi_n\rangle$ and $E_n$ are eigenvectors and eigenvalues of the Hamiltonian,
and the prime indicates that the ground state is excluded from the sum.

Our key result follows from the following simple consideration:
using the terminology defined above, we can express Eq. (\ref{1}) as
\begin{equation}
{\cal H}_l = {1 \over J_l}(H - J_r{\cal H}_r),
\label{4}
\end{equation}
which can be substituted into Eq. (\ref{2}) to give
\begin{equation}
{\cal H}_R(\theta) = J^\prime_r(\cos^2\theta-{J_r J^\prime_l \over J_l 
J^\prime_r}\sin^2\theta){\cal H}_r + {J^\prime_l \sin^2\theta \over J_l}H.
\label{5}
\end{equation}
The second term of the sum is a multiple of the Hamiltonian, and thus cannot
contribute to the scattering. The first term is proportional to ${\cal H}_r$ 
for all angles. Thus, within this theory, the observed two-magnon scattering 
spectrum will have the same line shape and peak position for all angles. 
This result is in direct contradiction with the arguments of Sugai.

The intensity of the spectrum is given by the expression
\begin{equation}
I(\omega,\theta) = (\cos^2\theta-{J_r J^\prime_l \over J_l J^\prime_r}
\sin^2\theta)^2I(\omega,0).
\label{6}
\end{equation}
This variation of angle can be used to determine the ratio of exchange
constants, provided one either knows $J^\prime_r/J^\prime_l$ or can relate 
it to $J_r/J_l$. This point is discussed a little later. In principle, one can 
perform the experiments by varying $\theta$ continuously, to obtain the above 
variation. One simply needs to keep the polarization direction of incoming and 
outgoing light fixed parallel to each other in the plane and rotate the sample.
To find the ratio of exchange constants, it is sufficient to consider
two angles:
\begin{equation}
{J_r \over J_l} = {J^\prime_r \over J^\prime_l}\sqrt{I(\omega,\pi/2) 
\over I(\omega,0)}.
\label{7}
\end{equation}
A simple procedure for determining the ratio of coupling constants 
could be to measure the maximum two-magnon Raman scattering intensity, 
then rotate the material through an angle of 90$^\circ$ and measure the
intensity again. The ratio of intensities would give the ratio 
of exchange constants, provided that one can reasonably estimate the value of 
the ratio of ${J^\prime_r / J^\prime_l}$.

\section{Analysis of Experimental Results}
\label{sec:analysis}

Experimental evidence does support the idea that the spectra do not vary much in
shape as a function of angle, but do vary in intensity. For example, Sugai and
Suzuki \cite{Sugai1} measured the Raman spectra for the spin-1/2 two-leg ladder 
materials La$_6$Ca$_8$Cu$_{24}$O$_{41}$ and Sr$_{14}$Cu$_{24}$O$_{41}$ in two
different configurations separated by an angle of 90$^\circ$. One configuration
had the incident and scattered polarizations parallel to the legs of the ladder, 
and the other had them parallel to the rungs. The peak of the observed spectra 
for the two configurations of La$_6$Ca$_8$Cu$_{24}$O$_{41}$ were found at 
3004 cm$^{-1}$ and 2948 cm$^{-1}$, a relative difference of less than 2\% which 
can be neglected. The spectral shapes are largely identical as well, but the
intensities are vastly different. The ratio of the intensities of the peaks of
these two spectra is approximately 0.52, which is quite significant. The same
measurements performed on Sr$_{14}$Cu$_{24}$O$_{41}$ yielded peaks at 3006 
cm$^{-1}$ and 3004 cm$^{-1}$, and an intensity ratio of approximately 0.39. The
peaks are even closer together than the others studied, and the intensity ratio
is smaller. Popovi\'c and collaborators \cite{Popovic} present scattering data
for Sr$_{14}$Cu$_{24}$O$_{41}$ that shows a similar lack of peak location shift,
as do measurements by Sugai and collaborators \cite{Sugai2} on LaCuO$_{2.5}$.
The small observed shift in the spectral peak could be due to impurities,
phonons, inter-ladder couplings as well as due to other intra-ladder
interactions not included here. \cite{Freitas&Singh} Note that because our
result derives from an operator relation, it is not sensitive to long range
ordering in the system, and should be valid for small interladder couplings.

There are several possiblities for the value of ${J^\prime_r / J^\prime_l}$ that
should be considered. We shall begin by examining the possibility that 
${J^\prime_r / J^\prime_l} = {J_r / J_l}$, as follows from a large-U
perturbative treatment and suggested by the work of Moriya
\cite{Moriya} and Shastry and Shraiman. \cite{Shastry&Shraiman} If we assume 
that ${J^\prime_r / J^\prime_l} = {J_r / J_l}$, Eq. (\ref{6}) becomes 
$I(\omega,\theta) = (\cos{2\theta})^2I(\omega,0)$. Thus, all of the maxima
should be identical in magnitude. This conclusion is directly contradicted by 
all of the experimental results discussed previously, where the ratio was
found to vary by more than a factor of 2. Clearly, direct 
proportionality does not exist.

A second possibility is that all of the 
nearest-neighbor Raman operator coupling constants are
equal to one another. This has been assumed
universally by almost all previous studies
of these ladder materials. The bonds along the rungs and legs of the
ladder are nearly identical, which is perhaps what led to this
assumption. However, it leaves open the question of why this ratio
will remain unity when the ratio $J_r/J_l$ deviates
significantly from unity. Assuming $J^\prime_r / J^\prime_l \approx 
1$, Eq. (\ref{7}) becomes 
\begin{equation}
{J_r \over J_l} = \sqrt{I(\omega,\pi/2) \over I(\omega,0)}. 
\label{7a}
\end{equation}
A third possibility can be motivated by perturbation theory. 
If the two-magnon Raman process involves a direct exchange
it is a second order process, whereas the usual
superexchange could be mediated by non-magnetic intermediate
ions and thus could be a fourth order process in a large U expansion.
In this case, a more natural relationship  between
the Raman and Heisenberg coupling constants is $J^\prime \propto \sqrt{J}$. So,
\begin{equation}
{J_r \over J_l} = {I(\omega,\pi/2) \over I(\omega,0)}. 
\label{7b}
\end{equation}
Using these two possibilities, we computed the ratio of the Heisenberg coupling 
constants for the materials studied by Sugai and his collaborators, using his 
published data. The results for these materials are shown in Table
\ref{theTable}.

Let us discuss these results in light of 
previous studies \cite{Dagotto&Rice,Dagotto}.
Some authors have adopted the point of view that this ratio
is close to unity, and have used that as the starting point 
of their analysis \cite{Sandvik&Scalapino}.
On the other hand, some local density approximation 
calculations \cite{Arai,Muller} find the ratio to be closer to $0.5$.
A number of other studies which allow the ratio to vary, also find values
close to $\sim 0.5$ \cite{Johnson,Greven&Birgeneau,Imai}, although
a range of experimental values from 0.5 to 1.13 have been quoted
\cite{Imai,Kumagai,Carretta,Takigawa,Magishi}. 
Brehmer
{\it et al.} \cite{Mikeska} make an interesting attempt to reconcile the 
conflicting viewpoints, keeping the ratio to be unity but allowing
an additional biquadratic ring-interaction in the Hamiltonian.

In the Sugai and Suzuki paper, the authors used the incorrect
argument for the energy shift to estimate the ratio of exchange couplings along 
the rungs and the legs of the ladder. Their ratio was determined to be 0.95 for 
La$_6$Ca$_8$Cu$_{24}$O$_{41}$ and 1 for Sr$_{14}$Cu$_{24}$O$_{41}$. 
It is now clear that Sugai and collaborators obtained values close to unity 
because, to a good approximation, the spectra do not shift at all with change in
polarization direction. Using their data and Eq. (\ref{7a}) and Eq. (\ref{7b}), 
we calculate the ratio to be less than one in all cases. It is 
interesting to note that using Eq. (\ref{7b}), our calculated ratio agrees
well with the conclusion that $J_r/J_l \sim 0.5$.

\section{Conclusion}
\label{sec:conclusion}

We have seen how Fleury-Loudon-Elliott theory predicts that the shape of
the Raman spectra in the spin-ladder geometry, with nearest-neighbor
Heisenberg exchange constants, does not change with the polarization
directions. Instead, there exists a relationship
between the intensity of two-magnon Raman scattering 
in different polarizations and the ratio of Heisenberg exchange
constants.
With some suitable assumptions about the Raman hamiltonian,
we have used it to estimate the ratio of rung to leg exchange constant
in several cuprate materials. The full dependence of the intensity
on the polarization direction can be experimentally verified and
should serve as a test for the Fleury-Loudon-Eliott theory.

\acknowledgements

The authors would like to thank T. Imai and A. Slepoy for their helpful 
comments. This work was supported in part by the National Science Foundation 
under grant number DMR-9986948.


\pagebreak
\begin{table}
\caption{Computed ${J_r / J_l}$ for the materials studied by Sugai 
{\it et al.} for the two ratios of the Raman coupling constants discussed in 
the text.}
\label{theTable}
\begin{tabular}{lcr}
Material&$J^\prime_r / J^\prime_l \approx 1$&$J^\prime_r / J^\prime_l = 
\sqrt{J_r / J_l}$\\
\tableline
La$_6$Ca$_8$Cu$_{24}$O$_{41}$&0.72&0.52\\
Sr$_{14}$Cu$_{24}$O$_{41}$&0.63&0.39\\
LaCuO$_{2.5}$&0.74&0.55\\
\end{tabular}
\end{table}

\end{document}